 \documentclass[referee]{aa} % for a referee version
\usepackage{graphicx}

\begin{document}
\title{Galactic structure studies with BATC star counts}

\author{Cuihua Du\inst{1},
        Xu Zhou\inst{1},
        Jun Ma\inst{1},
        Alfred Bing-Chih Chen\inst{2},
        Yanbin Yang\inst{1},  
        Jiuli Li\inst{1},
        Hong Wu\inst{1},
        Zhaoji Jiang\inst{1},
	Jiansheng Chen\inst{1}
	         }

\offprints{Cuihua Du}

\institute{National Astronomical Observatories, Chinese
              Academy of Sciences, Beijing 100012, P. R. China\\
              \email{dch@vega.bac.pku.edu.cn}.
              \and Department of Physics, National Cheng Kung University, 
           Taiwan 70148, Taiwan
            }

\date{Received 16 January 2003/ Accepted 21 March 2003}
\abstract{
We report the first results of star counts carried out with the
National Astronomical Observatories (NAOC)
$60/90$ cm Schmidt Telescope in
15 intermediate-band filters from 3000 to 10000 {\AA} in the BATC survey. 
We analyze a sample of over 1400 main sequence stars ($14\le$V$\le21$),
which lie in the field with central coordinates ${\rm
R.A.=09^h53^m13^s{\mbox{}\hspace{-0.13cm}.}30}$ and
DEC=47$^\circ49^{\prime}00^{\prime\prime}{\mbox{}\hspace{-0.15cm}.0}$ (J2000)
(Galactic coordinates: $l=169.95^{\circ}, b=49.80^{\circ}$). 
The field of view is 0.95 deg$^{2}$, and the spatial scale 
was  $1\arcsec{\mbox{}\hspace{-0.15cm}.}67$.  
Since star counts at high galactic latitudes are not strongly
related to the radial distribution, 
they are well suited to study the vertical distribution of the Galaxy.
In our model, the distribution of stars perpendicular 
to the plane of the Galaxy is given by two 
exponential disks (thin disk plus thick disk) and a de Vaucouleurs halo.
Also, based on star counts, we derive 
the scale heights 
of the thin disk to be $320^{+14}_{-15}$ pc and of the thick disk 
to be $640^{+30}_{-32}$ pc, respectively, with a local density of $7.0\pm1\%$ 
of the thin disk. 
The errors of scale heights and the corresponding space number 
density normalization are estimated at a $68\%$ confidential level.
The density law for the Galactic halo population is also investigated.
We find that the observed counts support an axial ratio of $c/a\le0.6$ for a 
de Vaucouleurs $r^{1/4}$ law, implying a more flattened halo.
 We consider that it is possible that the halo has two subpopulations--a flattened 
inner halo and a spherical outer halo in the Milky Way, and such a halo model might
resolve many of the divergences in star count results.
We also derive the stellar luminosity function (SLF) for the thin disk,
and it partly agrees with the Hipparcos luminosity function.

 }

\authorrunning{Cuihua Du et al.}

 \maketitle

%___________________________________________________________________
%Section 1
\section{Introduction}
Over the past decade considerable efforts 
have been undertaken to gain information
about the structure of the Galaxy. For example, 
M\'{e}ndez (1996) has developed a Galactic structure
model to predict star-counts self-consistently with kinematic 
distribution of stars; 
For example, novel multivariate data analysis methods are being applied to recent surveys
to characterize the individual stellar distributions of each populations
(Chen 1997; 1999); 
age-metallicity or the kinematics of nearby halo and disk stars have been 
determined (Norris 1998), and so on. 
The starcount method which is predominantly  
used to 
study the general properties of the Galaxy is a very effective way of 
constraining global structural parameters for the Milky Way, and of providing
density distribution of the Galaxy components. 

Bahcall \& Soneira (1980) established the first self-consistent model 
by analyzing star count data and derived that
the old thin disk has a scale height 
of 325 pc, and that the halo can be represented by a deprojected de Vaucouleurs 
$r^{1/4}$ law. Bahcall \& Soneira (1984; 1985) showed that their two major 
components (thin-disk and spheroid) model for the distribution of stars in the Galaxy is 
in agreement 
with all the available data. However,
Gilmore (1984) pointed out that the star count data cannot be fitted much 
better by a two component model, and proposed a Galaxy model with three populations: 
the halo, thin disk and thick disk.  At the same time, 
the scale height of the thick
disk was derived by Gilmore (1984) to be 1450 pc with a space number density 2\% of the thin 
disk in the solar neighborhood. Ojha et al. (1996) found that 
the vertical distribution of stars fainter than $M_{v}=3.5$  has scale height
$ h_{z}=260\pm50$ pc in the thin disk, and $ h_{z}=760\pm50$ pc in the thick disk with a local
density of 7.4\% relative to the thin disk. Chen (2001) used high latitude 
star counts to derive  the scale height
of the old thin disk to be $330\pm3$ pc, and the thick 
disk between 580 pc and 750 pc with a local density of 13\%-6.5\% of the thin disk
in their best-fit model.

For a spherical halo, the standard references 
for star counts are Kron (1980),
Koo \& Kron (1982; 1986) for two fields: SA 57 (near the North Galactic pole)
and SA 68 (near $l=90^{\circ}, b=-46^{\circ}$). Using a de 
Vaucouleurs $r^{1/4}$ law spheroid, 
Bahcall \& Soneira (1984) found that this axial ratio
was at least 0.8. However, Wyse \& Gilmore (1989) argued on the basis
of star counts that the axial ratio for the stellar halo is much flatter 
than this canonical value. They made an independent analysis of two fields,
the south Galactic Pole and one at $l=272^{\circ}, b=-44^{\circ}$, 
and obtained a very different ratio of 0.6, implying a more flattened halo.
Based on star counts
from the APS catalog of the POSS, 
Larsen (1994) gave a mean ratio of 0.5 implying $c/a\leq0.5$ for the stellar 
halo.

Although the existence of both the disk and halo
components seems well established, the spatial distribution of 
the Galactic components are not well determined and remain
controversial due to different and conflicting results from 
modeling of star counts. This is unfortunate because the shape of
Galactic components are necessary for understanding the conditions during the
formation of the Galaxy. 
In addition, the stellar luminosity function 
of main sequence stars in the disk is also a topic of debate. 
For example, there are some arguments about
whether features like the Wielen dip near $M_v\sim 7$ mag and the apparent 
rise of the stellar luminosity function  beyond $M_v\sim 12$ exist in the local neighborhood.
Therefore, 
it is worthwhile to know 
if the features are present in the BATC starcount.
 
To pin down the Galactic structure with more precision, 
the Beijing-Arizona-Taiwan-Connecticut (BATC) multicolor photometric
survey provides new catalogs with accurate object classification.
These catalogs are very useful
in constraining the structure of the main components of the Galaxy.
In this paper,
based on the BATC observation, we investigate the vertical distribution of
stars in the Milky way. Details of observations and data
reduction are given in Sect. 2; Sect. 3 describes the object classification
and the photometric parallaxes; Sect. 4 deals with the space density distribution
of stars;  Sect. 5 gives the SLF of the thin disk. 
Finally, we summarize our main conclusions and briefly mention the prospects for future work on
this survey project. 
%___________________________________________________________________
%section 2
\section{BATC Observations and data reduction } 
\subsection{BATC photometric system and observations}
The BATC survey performs photometric observations with a large field
multicolor system. 
There are 15 intermediate-band filters in the BATC filter system, which 
covers an optical wavelength range from 3000 to 10000 {\AA}  
(Fan et al. 1996; Yan et al. 2000; Zhou 2001).
The observation is carried
out with the
60/90 cm f/3 Schmidt Telescope of National Astronomical
Observatories (NAOC), 
located at the Xinglong station with an altitude of 900 m.
A Ford Aerospace 2048$\times$2048 CCD
camera with a 15 $\mu$m pixel size is mounted at the main focus of the
Schmidt telescope. The field of view of the CCD is $58^{\prime}$ $\times $ $
58^{\prime}$ with a pixel scale of $1\arcsec{\mbox{}\hspace{-0.15cm}.} 67$.  
 
As in the definition of the AB system of Oke \& Gunn (1983), the  BATC magnitude
system is defined as follows:  
\begin{equation}
 m_{\rm batc} = -2.5\cdot {\rm log}\widetilde{F_{\nu}} - 48.60, 
\end{equation}
where $\widetilde{F_{\nu}}$ is the flux per unit frequency in unit of
$\rm erg~s^{-1}cm^{-2}Hz^{-1}$.
The advantage of the AB magnitude system
is that the magnitude is directly related to physical units. 
The 4 Oke \& Gunn (1983) standards are used for flux calibration in the 
BATC survey. The 4 stars are HD19445, HD84937, BD+262606
and BD+174708. The magnitudes of these standards were refined by several
authors. Fukugita et al. (1996) presented the latest re-calibrated
fluxes. Their magnitudes were also corrected
in the BATC photometric system (Zhou et al. 2001).
The extinction coefficients and magnitude zero points
obtained from standard star observations are then used to calibrate 
other BATC field 
images. The observation for calibration are described in detail 
in Zhou et al. (2001).

The observations of the BATC T329 field 
(Galactic coordinates: $l=169.95^{\circ}, b=49.80^{\circ}$) were obtained
in 15 intermediate band
filters with a total exposure time of 69.37 
hours from December 11, 1996 to May 5, 1999. 
The CCD images are centered at ${\rm
RA=09^h53^m13^s{\mbox{}\hspace{-0.13cm}.}30}$ and
DEC=47$^\circ49^{\prime}00^{\prime\prime}$ (J2000). 
The resulting images with the total exposure
time and image numbers of each filter band are listed in Table 1. 

% Table 1_______________________________
   \begin{table}
     \caption[]{Parameters of the BATC filters and statistics of observations 
for the BATC T329 sample (from 1996 to 1999)}
         \begin{tabular}{cccccccc}\hline
            \hline
           No. & Filter &  Wavelength  & FWHM &Total Exp. &
           Number of & Number of & Calib. \\
                &         & (\AA)   & (\AA)   & (hour)   &
           images & calibrations & error\\

%           No. & Filt & $\lambda\lambda$ & Exp. &
%           Num.of & Num.of & Calib. \\
%                &         & (\AA)      & (hour)    &
%           images & Calib. & error\\
           \hline

1  & $a$ & 3371.5  & 359   & 03:00& 06 &1 & 0.088 \\ 
2  & $b$ & 3906.9  & 291   & 03:40& 11 &1 & 0.023 \\
3  & $c$ & 4193.5  & 309   & 02:40& 05 &5 & 0.007 \\
4  & $d$ & 4540.0  & 332   & 07:00& 21 &5 & 0.008 \\
5  & $e$ & 4925.0  & 374   & 06:10& 16 &8 & 0.006 \\
6  & $f$ & 5266.8  & 344   & 05:30& 17 &6 & 0.003 \\
7  & $g$ & 5789.9  & 289   & 05:10& 16 &6 & 0.005 \\
8  & $h$ & 6073.9  & 308   & 04:00& 13 &3 & 0.007 \\
9  & $i$ & 6655.9  & 491   & 05:10& 20 &8 & 0.009 \\
10 & $j$ & 7057.4  & 238   & 06:20& 21 &3 & 0.009 \\
11 & $k$ & 7546.3  & 192   & 03:50& 12 &5 & 0.002 \\
12 & $m$ & 8023.2  & 255   & 06:00& 16 &2 & 0.009 \\
13 & $n$ & 8484.3  & 167   & 04:20& 09 &3 & 0.006 \\
14 & $o$ & 9182.2  & 247   & 04:32& 11 &2 & 0.005 \\
15 & $p$ & 9738.5  & 275   & 01:00& 10 &2 & 0.004 \\

 \hline
         \end{tabular}
\\ 
\\
{\fontsize{8}{12} NOTES.--- Listed are the Parameters of the BATC filters and 
statistics of our observation sample. Col. (1) and col. (2) represent
the ID of the BATC filters. Col. (3) and col. (4) represent the central wavelengths
and FWHM of the 15 BATC filters, respectively. From col. (5) to col. (7), 
we list the the total exposure time, number of images and number of calibration 
of each filter band. The last column gives the calibration error in magnitude.}
   \end{table}

%%%%%%%%%%%%%%%%%%%%%%%%%%%%%%%%%%%%%%%%%%%%%%%%%%%%%%%%%%%%%%%%%%%%%%
%Figure 1---------------------------------------
\begin{figure}
\includegraphics[angle=-90,width=140mm]{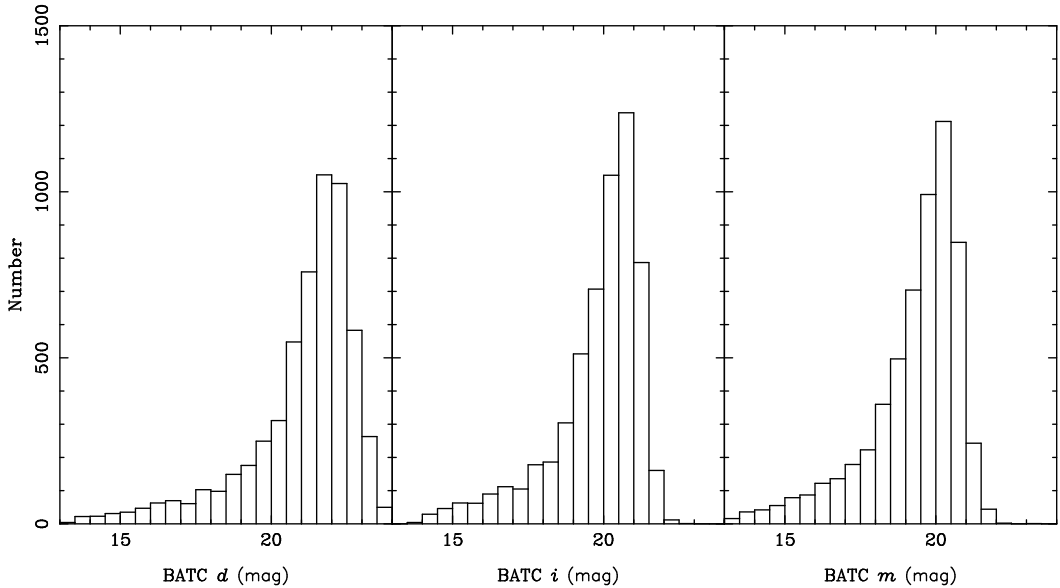}
\caption{The distribution of the BATC $d$, $i$, $m$ filter for 6000 objects 
in the BATC T329 field}
\end{figure}
\subsection{Data reduction}
The BATC survey images were reduced through standard procedures, 
including bias subtraction, flat-fielding correction and 
flux calibrations (Fan et al. 1996; Zhou et al. 2001;
Zhou et al. 2003). After the basic correction
described above, the flat-fielded images and field images
observed of each filter were combined by integer pixel shifting, 
respectively. 
When combining the images, the cosmic rays and bad pixels were corrected
by comparison of multiple images.
The HST Guide star catalog (GSC) (Jenkner et al. 1990) was then 
used for coordinate determination.
The final RMS error in positions of GSC stars
is about 0.5 arcsec. 

The magnitude of the point sources in the BATC fields are measured 
by the photometric
method of point spread function (PSF) fitting. 
Our PSF magnitudes were obtained through an automatic data reduction 
program PIPELINE 2, which  was developed based on Stetson's DAOPHOT 
procedures (Stetson 1987).

It is difficult to obtain good PSF fitting for faint stars
due to their asymmetry in profile. However, in order to 
construct a SED catalog as complete as possible, faint 
stars were measured with aperture photometry. 
The difference between the PSF and aperture
photometry result is corrected by the bright, isolated stars
in the same field.  
At the completion of photometry, the SEDs of all measurable
objects are obtained. Finally,
we determine the SEDs of 15 filters for more than 6000 objects
in the BATC T329 field. 
In Fig. 1, we plot the BATC  $d$, $i$, $m$  magnitude histograms for the
6000 objects in the field. 
In general,
the limiting magnitude of our photometry is about
$21\hspace{0.1cm}{.}\hspace{-0.15cm}^{m}0$ with an error of about
$0\hspace{0.1cm}{.}\hspace{-0.15cm}^{m}1$ in the BATC $i$ band.

\section{Object classification and photometric parallaxes}
One of the goals of the BATC survey is the classification of every 
object found in all BATC fields, and the majority of these objects
should be classified according to their SED information constructed from
the 15-color photometric catalog. The observed colors of each object
are compared with a color library of known objects with the same
photometric system. 
For each object the probability of belonging to a certain object class (stars-galaxies)
is given. 
Following that, we use spectral templates of galaxies and stars to 
discriminate the stars from galaxies. 
The profile of objects classified as stars does not deviate significantly from 
that of stellar templates.
The input spectral templates for the galaxies
are the enlarged galaxy evolutionary library of Bruzual and Charlot (1993), and 
the empirical template that is obtained through the averaged spectra of observed local field
galaxies (Coleman 1980). Details about the classification of galaxies are given in
Xia et al. (2002). 
The input library for stellar spectra is
the Pickles (1998) catalog. This library consists of 131 flux-calibrated spectra, including 
all normal spectral types and luminosity classes at solar abundance, and metal-weak and metal-rich
F$-$G dwarfs and G$-$K giant components. Our sample may contain stars spread over 
a range of different metallicities. In Fig. 2, the two-color diagram $(d-i)$ versus $(i-m)$
is shown for our sample. Although the 15 filters are used in the object classification,
the two-color diagram based only on the $d$, $i$, $m$ filters as
an example shows that the scatter still exists in our sample.
Most of stars lie in the mean main sequence track, and for those objects beyond the track,
the scatter can be mainly due to the metallicity effect.
Secondly, the analysis of deep star counts can be critically affected by faint galaxy contamination.
At the magnitude limit of our data ($i\sim 21$), galaxies outnumber stars.
Although our multicolor photometric information is effective at discriminating stars from galaxies,
 some fainter objects still can be misclassified due to the increase in 
the observational error associated with the 
photometry of faint stars. The possibility of galaxy contamination is estimated to be less than 3\%. 
In addition, there is an effect called `non-simultaneous observation', 
and it will play an important role for long period variable stars (Chen 1996). However, this should not
affect the final results derived by our sample, since the variable stars are relatively few. Here, we
can only give the upper limit (less than 2\%)(Zhang 2002; private communication).

For those stars, the probability of belonging to
a certain star class is computed by the SED fitting method. The 
standard $\chi^2$ minimization, i.e., computing and minimizing the deviations between photometric
SED of a star and the template SEDs obtained with the same photometric system, 
is used in the fitting 
process. The minimum $\chi^2_{min}$ indicates the best fit to the observed SED by
the set of template spectra:
\begin{equation}
{\chi^{2}=\sum\limits_{l=1}^{N_{filt}=15}\left [\frac{F_{obs,l}-
 F_{temp,l}-b} {\sigma_{l}} \right ]^{2}},
\end{equation}
where ${F_{obs,l}}$, $F_{temp,l}$ and $\sigma_{l}$ are the observed fluxes,
template fluxes and their uncertainty in filter $l$, respectively, and
$N_{filt}$ is the total number of
filters in the photometry, while $b$ is the mean magnitude difference between 
observed fluxes and template fluxes.

Thus, we can 
obtain the spectral types and luminosity classes for stars 
in the BATC survey. Due to the multicolor photometry, 
the classification can be relatively accurate. 
After knowing the stellar type, the 
photometric parallaxes can be derived  by estimating 
absolute stellar magnitudes. We adopted
the absolute magnitude versus stellar type relation for main-sequence stars from Lang (1992). 
The derived relation $M_{v}$ vs. $(d-i)$ for the main sequence is shown in Fig. 3.
In this Figure, the scatter band in the relation $M_{v}$ vs. $(d-i)$
is probably due to the larger uncertainty in the absolute magnitude determination 
only using two colors to express the information resulting from 15 colors.
 
A variety of errors affect the determination of stellar distances. The first source
of errors could be from photometric uncertainty; the second from the misclassification 
that affects the derivation
of absolute magnitude, but the  misclassification should be small 
due to the multicolor photometry. For luminosity class V, types F--G--K, the absolute 
magnitude uncertainty is about 0.3 mag, and 0.8 or more for late M main sequence stars.
In addition, there may be an error by the contamination of 
binary stars in our sample. We neglect the effect of binary contamination on distance 
derivation. For binaries with equal mass components, the distance will be assumed 
closer by a factor of $\sqrt{2}$. Due to the unknown but probable mass distribution in binary 
components, the effect is certainly less severe (Kroupa 1993; Ojha 1996). 
Since the BATC T329 field is at intermediate 
latitude ($b=49.80^{\circ}$), we can neglect the influence of interstellar 
extinction in distance calculation. 

%Figure 2________________________________________________
\begin{figure}
\includegraphics[angle=-90,width=80mm]{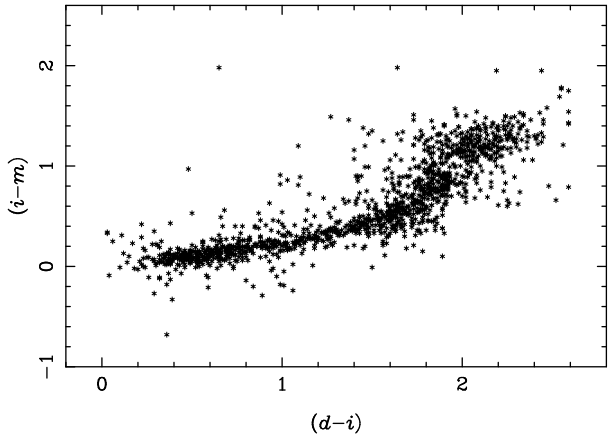}
\caption{The distribution of ($i-m$) versus ($d-i$) for
the BATC T329 field down to the limiting magnitude, $i\le21$.}
\end{figure}
%Figure 3________________________________________________
\begin{figure}
\includegraphics[angle=-90,width=80mm]{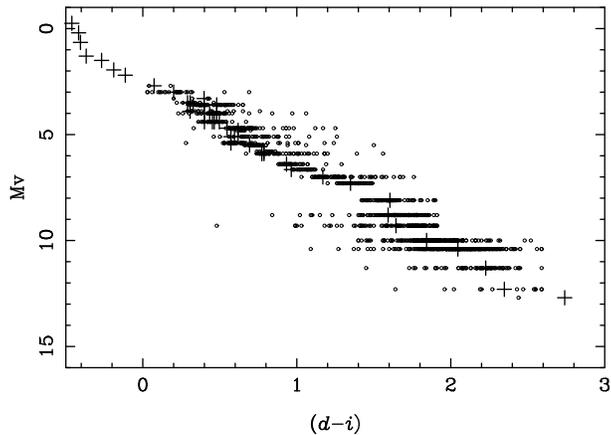}
\caption{$M_{v}-$($d-i$) CMD for the main sequence stars is shown.The open 
circles refer to stars of the BATC T329 field, and the crosses refer to stellar
library of Pickles (1998).}
\end{figure}

\section{Space density distribution}
In previous work, various models have been developed
to describe the stellar populations of the Galaxy. 
In general, these models
were based on the assumption of a suitable
spatial density distribution, and on
the observational luminosity function and color-magnitude diagram for each
stellar population (Bahcall \& Soneira 1984; Reid \& Majewski 1993) to fit the 
structural parameters by exploiting the measurements of colors and magnitudes.
Here, on account of the use of the photometric parallaxes, we can make a direct evaluation
of the spatial density law. Rather than trying to fit the structure of the Galaxy in the observed
parameter space of colors and magnitudes, we translate the observations into discrete density 
measurements at various points in the Galaxy.

As the starting point for our analysis, we consider models with 
three major components: a thin disk and a thick
disk, both with exponential density profiles, and a halo component,
with a deprojected de Vaucouleurs profile.
Given a reliable stellar sample, the next step is to differentiate between
member stars of the disk and those of the halo population.
It is important to consider the volume-sampling effects inherent in any
star count analysis. As described in Reid et al. (1993; 1996),
the convolution of a monotonically-decreasing density law with the sampling
volume leads to a `preferred' distance modulus
dependent on the slope of the density law along the
particular Galactic line of sight for a given
stellar population.  Since the density distribution of
the disk population falls off faster with height above the plane than 
does that of the halo, and the former is sampled primarily at 
$m-M\sim7$ mag, while the latter lies preferentially at $m-M\sim17$  
mag. Thus, at a given apparent magnitude, 
the halo contributes more luminous (bluer) stars
than the disk.
In Fig. 4, we plot the color distribution of our sample stars. 
From this figure, we can see the bimodal color distribution and the left
peak is a close to a normal distribution. 
We use a Gaussian function to fit it and check the convergence 
by its left wing which is not contaminated by disk stars.
We find that the bluer stars are dominated by halo stars which are supposed to be metal-poor,
with a turnoff of $d-i\sim0.9$. Beyond this point the contribution
by disk stars becomes dominant.
Therefore, a crude disk-halo 
separation can be drawn by color cut.
%Figure 4________________________________________________
\begin{figure}
   \includegraphics[angle=-90,width=80mm]{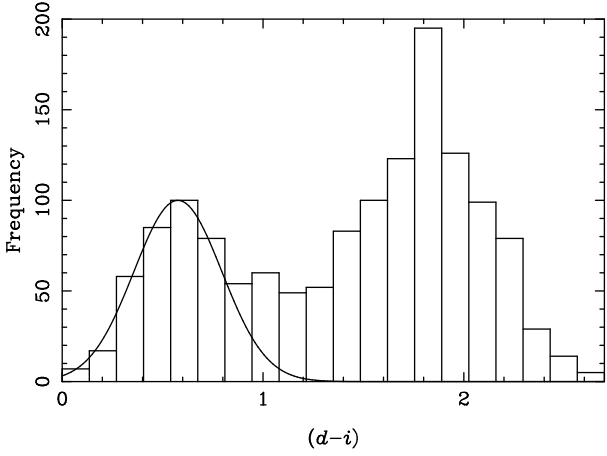}
   \caption{Distribution of ($d-i$) in the field, this bimodal distribution can 
be used to separate halo and disks by a color cut.}
   \end{figure}
%Figure 5________________________________________________
\begin{figure}
   \includegraphics[angle=-90,width=80mm]{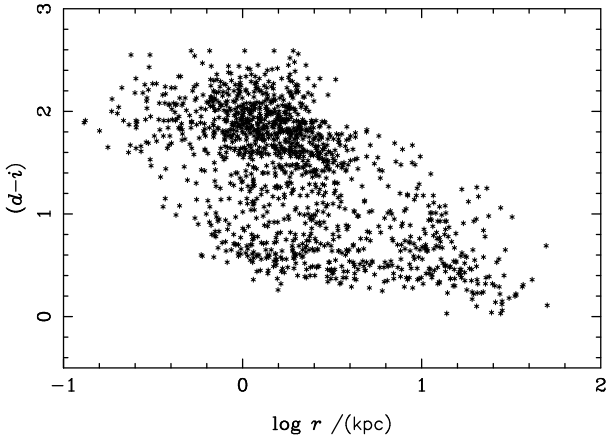}
   \caption{Spatial distribution of ($d-i$), and the photometric distances are derived
  according to the stellar type.}
   \end{figure}
We try to derive the structural parameters (e.g. scale height)
of the thin and thick disk populations using our data set at intermediate
latitude. For this we calculate the stellar space density
as a function of distance from the Galactic plane. At first, we use the two
dimensional distribution of stars in the ($d-i$) vs. log$~r$ diagram (Fig. 5)
to correct for this 
incompleteness. The nearest bins are assumed to be complete;
for the incomplete bins we multiply iteratively by 
a factor given by the ratio of complete to incomplete number counts
in the previous bin (Phleps et al. 2000).
With the corrected number
counts, the density in the log arithmetic space volume bins $V_j$ can then be
calculated according to
\begin{equation}
\rho_j=\frac{N_j^{corr}}{V_j}
\end{equation}
here, $V_j=(\pi/180)^2(\omega/3)(r_{j+1}^{3}-r_j^3)$ is 
partial volume, 
$r_{j+1}$ and $r_{j}$ are the limiting distances; and $\omega$ is field size in square degrees.

\subsection{Density distribution in the disk}
We study the density distribution of the disk stars by taking stars
of $(d-i)\ge0.9$. The color cut is a crude separation
between disk and halo, but we can gain a clear insight to
the disk distribution due to the relatively less contamination by halo
stars.  Fig. 6 shows the resulting density
distribution of the disk stars in our field. The solid line represents
a fit with a superposition of two exponentials, i.e.,
the function of the disk is:
\begin{equation}
\rho(z)=n_{1}exp(-z/h_{1})+n_{2}exp(-z/h_{2}),
\end{equation}
$h_{1}$ and $h_{2}$ are the scale heights of 
the thin disk and thick disk, respectively,
and the dashed line is the fit for the thin disk component. 
It is obvious that a single exponential
disk is not a good fit for the disk component. Thus, the thick disk 
component in our Galactic model is indispensable to explain 
the observed star counts and color data. 
The comparison between data and simulations is made using a $\chi^2$-fit.  
The corresponding parameters can be given.
The scale height of the thin disk
$h_{1}$ is $320^{+14}_{-15}$ pc.  The thick disk's 
scale height $h_{2}$ is  $640^{+30}_{-32}$ pc, 
and the corresponding space number density normalization is $7.0\pm1\%$
of the thin disk.
The errors of scale heights and the corresponding space number 
density normalization are estimated at a $68\%$ confidential level.

%Figure 6________________________________________________
   \begin{figure}
   \includegraphics[angle=-90,width=80mm]{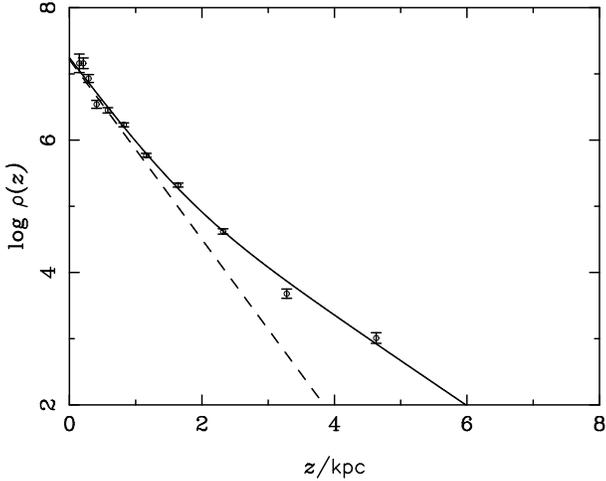}
   \caption{Vertical density distribution of the disk stars ($d-i>0.9$) in the field. 
The solid line is a fit with a superposition of two exponentials, the dashed line is the 
single exponential fit for the thin disk component.}
   \end{figure}
%Figure 7________________________________________________
   \begin{figure}
   \includegraphics[angle=-90,width=80mm]{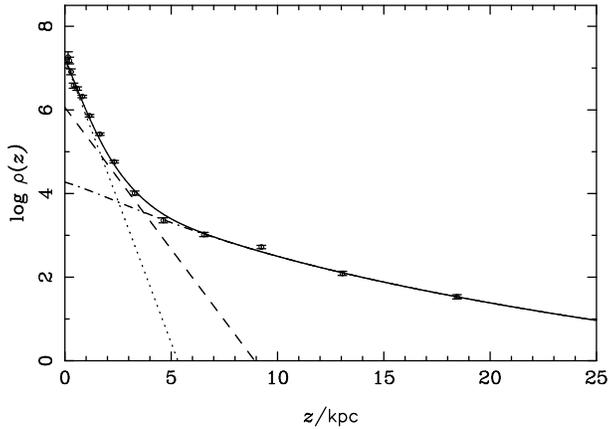}
   \caption{Density distribution perpendicular to the Galactic plane, 
the dotted line shows the contribution of the thin disk component, 
the dashed line is the contribution of the thick disk, 
the dot-dashed line is a de Vaucouleurs law and the solid line the sum of the three.}
   \end{figure}

Some results for the thin disk scale height have been published in the 
literature. Most
authors (Gilmore 1984; Bahcall \& Soneira 1984; Yoshii 1987; Reid \& Majewski
1993) derived a scale height of 325 pc for old-disk stars;
Chen (2001) used two large star count samples to derive the scale 
height of the thin disk to be 330 pc. However, Kuijken \& Gilmore
(1989) derived a scale height of 249 pc for the old-disk stars. 
Haywood (1994) showed, from an analysis of numerous
star counts towards the pole using his self-consistent evolutionary model,
that the thin disk scale height does not exceed 250 pc.
Ojha (1999) also found that
the scale height of the thin disk is 240 pc based on an analysis of two
star count samples. It is clear that our derived thin disk scale height (320 pc) 
is consistent with the value presented by Chen (2001) and the classical value 
in the literature.

The thick disk density law can be reasonably modeled either by a exponential
or by a density law close to sech$^{2}(z)$, and star counts are 
unable to distinguish
between these two hypothesis (Reylr\'{e} \& Robin 2001). The thick disk
vertical structure is generally described as  exponential with  scale  heights
varying between 480 pc 
and 1500 pc and local density between 1\% to 15\% 
relative to the thin disk.
Because of the small
proportion of the thick disk locally with regard to the thin disk,
it is difficult to derive a accurate  scale height and local density of 
the thick disk. In general,
any values of $h_{z}$ in the range $480-1500$ pc and of local density
in $1\%- 15\%$ turn out to be acceptable.
 
The thick disk's scale height is anticorrelated with its local density when
fitted simultaneously, and
a small scale height is obtained in combination with high local density,
while large scale height is associated with
low local density  
(Robin et al. 1996; Chen 2001). Robin et al. (1996) derived a scale height of 
$h_{z}=760\pm50$ pc with a local density of $5.6\pm1.0$\% relative to the 
thin disk,
and Ojha et al. (1999) presented a scale height of 790 pc with a local density
of 6.1\% of the thin disk from a photometry and proper-motion survey in 
the two directions at intermediate latitude. Gilmore (1984) presented 
a scale height
of 1300 pc and local normalization of 2\%, Kuijken \& Gilmore (1989) derived a scale 
height of 1000 pc and local normalization of 4\%. Chen (2001) gives a thick
disk scale height between 580 pc and 750 pc, with a local density of $13-6.5\%$ of the thin disk. 
Our study gives a scale height of  $640^{+30}_{-32}$ pc 
and the corresponding space number density normalization is $7.0\pm1\%$ of the thin disk. 
In Table 2,
we summarize the main parameters relating to the vertical distribution of 
stars in these models. 
It is clear that our results favor the thick disk scale height presented 
by Chen et al. (2001), well below the original proposal of Gilmore et al. (1984).

Fig. 7 gives the density distribution of all stars in the BATC T329 field.
The dotted line shows the contribution of the thin disk component; the dashed line is the 
contribution of the thick disk; the dot-dashed line is a de Vaucouleurs law, and 
the solid line is the sum 
of the three components. It shows that 
the thick  disk dominates star counts at distances between 1.5 and 4 kpc over
the galactic plane. Phleps (2000) derived  that the thick disk provides the 
predominant contribution from 1.5 kpc to 5.0 kpc.
It showed that our thick disk result determination is in agreement
with Phleps's (2000).
However, photometric counts are not accurate enough to 
estimate the distances of stars at the turnoff with an accuracy of even a
factor of two. 
It can also be seen that the
corresponding plots fit the distribution of the halo stars up to distance
of over 20 kpc above the Galactic plane.

% Table 2_______________________________
   \begin{table}
     \caption[]{A comparison of the parameters obtained from different sources}
         \begin{tabular}{c|c|cc|cccc}\hline
            \hline
           Component & Thin disk & Thick disk &Thick disk  & &  Halo & & \\
                   
                     &  $h_{1}$ (pc)&  $h_{2}$(pc) & Local normalization & de Vaucouleurs 
                & &power law &\\
                    &  & & &Local normalization &axis ratio &index(n)&axis ratio\\

           \hline

Bahcall(1984)& 325 & ...&...& 0.15\%& 0.8 & ...& ...\\
Gilmore(1984)& 325 & 1300& 2\% & 0.125\%& 0.8&...&...\\
Kuijken(1989)& 249&1000 &4\%&...&...&...&...\\
Ojha(1999)&240&790&6.1\%&...&...&...&...\\
Robin(2000)&...&750&5.6\%&...&...&2.44&0.76\\
Chen(2001)&330&$580-750$&13-6.5\%&0.125\%&...&2.5&0.55\\
Siegel(2002)&280&$700-1000$&$6-10$\%&...&...&2.5&0.6\\
Larsen(1994)&...&...&...&...&0.48&...&...\\
This work& 320&640&7.0\%&0.125\%&0.58&...&...\\

 \hline
         \end{tabular}
        
   \end{table}
%%%%%%%%%%%%%%%%%%%%%%%%%%%%%%%%%%%%%%%%%%%%%%%%%%%%%%%%%%%%%%%%%%%%%%
\subsection{The density law of the halo}

We use a de Vaucouleurs law for the halo component of the Galaxy.
The de Vaucouleurs
law is an empirical description of the density distribution of the galactic
halo. The analytic approximation is:
\begin{eqnarray}
\rho_{H}(z,b,l)&=&\nonumber \rho_{0}\frac{exp[-10.093(\frac{R}{R_\odot})^{1/4}+10.093]}
{(\frac{R}{R_\odot})^{(7/8)}}  \\
&\nonumber\times&1.25 \frac{exp[-10.093(\frac{R}{R_\odot})^{1/4}  
+10.093]}{(\frac{R}{R_\odot})^{(6/8)}}, ~~R<0.03R_{\odot} \\ 
&\nonumber\times&[1-0.08669/(R/R_\odot)^{1/4}], ~~R\geq0.03R_\odot   \\  
\end{eqnarray}
where $R=(x^2+z^2/\kappa^{2})^{1/2}$ is Galactocentric distance, 
$\kappa$ is the axis ratio,
$x=(R_{\odot}^2+d^2$cos$^2b-2R_{\odot}d$~cos$b$~cos$l)^{1/2}$, $z=d$~sin$b$;
$R_{\odot}=8$ kpc is the distance of the sun
from the Galactic center, 
$b$ and $l$ are the Galactic latitude and longitude;
the normalization factor $\rho_{0}$ is usually expressed as a percentage
of the local spatial density of stars. 

In our sample, disk stars greatly outnumber 
halo stars and it is therefore difficult to isolate halo stars,
and we use the blue tail of the 
distribution (Fig. 4) $d-i<0.9$ mag for our star counts
to distinguish the population of halo stars from the sample.   
Adopting a de Vaucouleurs $r^{1/4}$ law halo
and a local density normalization $\rho_{0}=0.125\%$, the $\chi^2$-curve
for the axis ratio parameter as derived from the 
BATC T329 field is shown in Fig. 8. 
Our counts imply that the axial ratio of the stellar halo approximates 0.6.
This ratio agrees with the star count results of
Larsen (1994) and is also consistent with a kinematic analysis, but it does not 
agree with the ratio from the Koo et al. (1986) surveys. 
The apparent discrepancy is probably due to the multi-component 
nature of the Galactic halo (Buser 1985).
Some studies of the kinematics and abundance of both field stars
and globular clusters show that the halo is better described as 
having two subpopulations--a flattened inner halo and a spherical 
outer halo (Siegel 2002). Additional support for  dual-halo 
models can be drawn from the apparent dichotomy in detailed chemical abundance
of halo stars (Nissen \& Schuster 1997). In a dual-halo model, nearby stars (Larsen 94;
Chen 2001; Siegel 2002; this work) are dominated by the flattened inner halo while
distant stars (Koo 1986; Bahcall 1984; Nissen 1997) are dominated by the round outer halo.
Such models may resolve
many of the disagreements in star count results.

Sources of uncertainty in our star counts include the effects of extinction
and possible contamination by thick disk stars. 
Chen (2001) found that the redding for the BATC T329 field, $E_{B-V}<0.011$
and $A_{v}/E_{B-V}$ for the bluest band $a$ is 5.46, therefore, it yields
$A_{v}$ (the BATC $a$ band) 0.06 and lower than our observation error 0.1, 
so we can neglect its effect.
A major uncertainty in the interpretation of our star counts is 
from a thick disk. 
However, the small sample prevents further conclusions about
halo structure. 
We will recheck these results with forthcoming BATC data in other
Galactic directions in the future.
%Figure 8________________________________________________
   \begin{figure}
   \includegraphics[angle=-90,width=80mm]{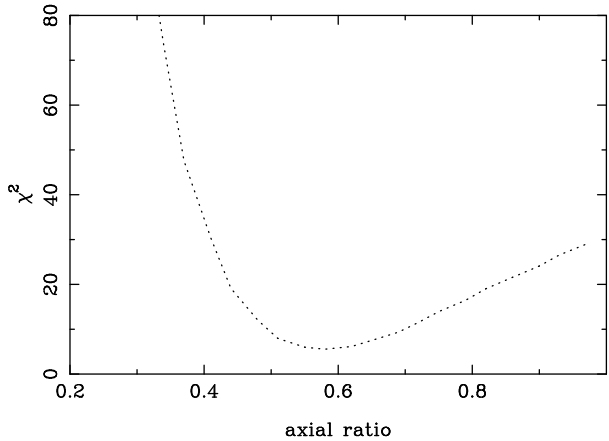}
   \caption{$\chi^2$-curve for the axis ratio of de Vaucouleurs $r^{1/4}$ halo,
derived from star count data of the BATC field.}
   \end{figure}

%Figure 9________________________________________________
\begin{figure}
   \includegraphics[angle=-90,width=80mm]{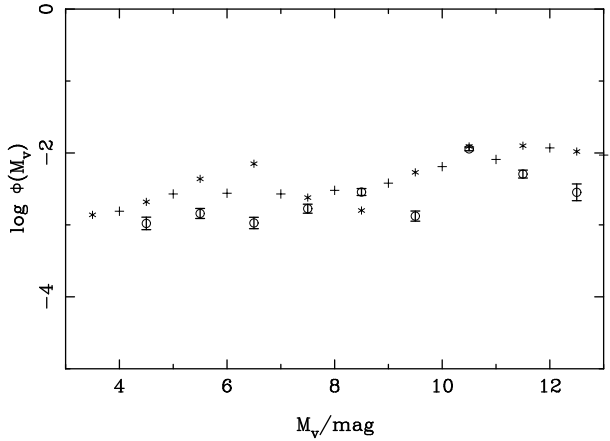}
   \caption{The SLF of the thin disk stars. The circles represent the present
determination, including error bars determined within the fit.
The crosses show the local SLF of Jahrei{\ss} (1997a), 
and the stars are derived from Eggen (1983).}
   \end{figure}

\section{The stellar luminosity function}
The stellar luminosity function (SLF) gives the number of stars per cubic
parsec of space per unit interval of absolute magnitude $M$ as a function 
of $M$. The SLF is a key input parameter for models of the stellar content
of the Galaxy. The stellar samples are used to derive it, and 
counts of stars are estimated to be complete at distance (from the sun) 
limits of 24 kpc and absolute magnitude limits of
12 mag. 
The stellar luminosity function differs in different parts of the Galaxy.
Of the greatest interest is its determination in the thin disk.
The stellar luminosity function for the thin disk stars can be calculated
according to the knowledge of the density distribution function. We selected
the nearby stars as thin-disk stars with distances less than 1.5 kpc.
Beyond this point the contribution from thick disk and halo stars 
becomes dominant. 
  
To investigate this possibility further, 
we make the following alternative estimate of the SLF. We first
calculate the effective volume, $\upsilon_{eff}(M_v)$, as a function
of absolute magnitude by integrating the volume element as a function
of the distance $r$ along the line of sight,  
\begin{equation}
\upsilon_{eff}(M_v)=\omega\int_{R_{min}}^{R_{max}}\nu(r,b)r^2dr,
\end{equation}
here, the integration limits are
\begin{displaymath}
R_{min}=10^{0.2(14-M_{V})-2.0}
\end{displaymath}
\begin{displaymath}
R_{max}=min(1.5~kpc, 10^{0.2(21-M_{V})-2.0}),
\end{displaymath}
and $\nu(r,b)$ is the distribution function of the thin disk, and it is normalized 
to unity at $z=0$.

Finally, we divide the total number of stars in a given magnitude bin by
the effective volume of that bin. 
The derived SLF is listed in Table 3 and in Fig. 9, in comparison
with SLF derived by Eggen (1983)
from a survey of proper motion selected stars with apparent magnitude
$V<15$, and the SLF by Jahrei{\ss} \& Wielen (1997) from the superior
HIPPARCOS data for the local stellar population. 

% Table 2_______________________________
   \begin{table}
     \caption[]{The logarithmic luminosity function for disk stars in the neighborhood, 
where $\upsilon_{eff}$ is effective volume in pc$^3$, $N$ is the number of stars in the absolute magnitude bin}
         \begin{tabular}{cccc}\hline
            \hline
         $ M_{V}$  & $\upsilon_{eff}$ & $N$ &  log~$\phi(M_V)$ \\
             
          \hline

(4, 5)    & 2.383(4)   & 25 & -2.98 \\
(5, 6)    & 2.706(4)   & 39 & -2.84 \\
(6, 7)    & 2.820(4)   & 30 & -2.97 \\
(7, 8) 	  & 2.855(4)   & 48 & -2.77 \\
(8, 9) 	  & 2.864(4)   & 82 & -2.54 \\
(9, 10)	  & 2.867(4)   & 38 & -2.88 \\
(10, 11)  & 2.387(4)   & 272 & -1.94 \\
(11, 12)  & 1.218(4)   & 62  & -2.29 \\
(12, 13)  & 4.935(3)   & 14  & -2.55 \\

 \hline
         \end{tabular}
   \end{table}

%%%%%%%%%%%%%%%%%%%%%%%%%%%%%%%%%%%%%%%%%%%%%%%%%%%%%%%%%%%%%%%%%%%%%%

As can be seen from Fig. 9, the observed luminosity function in the BATC T329 field is
in agreement with the local luminosity function as given by Eggen (1983) and
by Jahrei{\ss} \& Wielen (1997) only for three intervals, i.e.: $7<M_{v}<8$,
$8<M_{v}<9$, and $10<M_{v}<11$, whereas the observed SLF is lower than the Wielen (1997)
standard for the other intervals. The low value of the present observed SLF at the brighter
luminosity intervals is most probably due to the larger uncertainty in the absolute magnitude 
determination. It is noticeable that there is a dip at the faint luminosity 
intervals $11<M_{v}<12$ and $12<M_{v}<13$. It is possible that the dip is real at the 
faint magnitude end.
In addition, the discrepancy at the 
faint magnitude intervals may come from 
the fact that the stars are too faint to be sampled 
in any significant number and uncertainties of the 
classification may introduce systematic errors that 
are hard to quantify.

\section{Conclusions}
Based on the BATC observation, we have analyzed the star counts with 
the help of a Galaxy model
in order to parameterize the vertical distribution of stars 
in the Milky Way.  
The basic conclusions of this paper are:\\
1. Based on the BATC multicolor photometry and on the template SEDs, 
the stellar spectral types and luminosity classes can be
obtained. Thus, the photometric parallaxes 
of the main sequence stars can be derived.\\ 
2. Using two exponential disks, we determine that 
the scale height of the thin disk is $320^{+14}_{-15}$ pc and the thick 
disk scale height is $640^{+30}_{-32}$ pc with a corresponding 
space number density 
normalization  $7.0\pm1\%$ of the thin disk. \\
3. Adopting a de Vaucouleurs $r^{1/4}$ law halo
and a local density normalization $\rho_{0}=0.125\%$,
the observed counts yield a axial ratio of $c/a\le0.6$, 
implying a more flattened halo.
Our study suggests that it is possible that the halo has two subpopulations--a flattened 
inner halo and a spherical outer halo in the Milky Way.  \\
4. Based on this knowledge of the density distribution, we determined 
the stellar luminosity function (SLF) for the thin disk stars,
with distances less than 1.5 kpc. Beyond this point the contribution
by thick disk and halo stars become dominant. 

In general, these preliminary test results are in remarkable 
agreement with the majority of
independent recent determinations.
Since our data comprise a large field,
with a multicolor filter system including 15 intermediate-band filters,
this finding confirms in a satisfactory way similar results found from
other projections.  
Of course, the structural parameter values derived from the BATC survey
data critically depend on the reliability of the basic model.
While this model needs to be refined in many respects in the future, it can 
provide a suitable framework for interpreting the BATC data in its present form.

\begin{acknowledgements}
We would like to thank the anonymous referee for his/her
insightful comments and suggestions that improved this paper.
The BATC Survey is supported by the
Chinese Academy of Sciences, the Chinese National Natural Science
Foundation under the contract No. 10273012, the Chinese State 
Committee of Sciences and
Technology. This work has been
supported by the National Key Basic Research Science
Foundation (NKBRSF TG199075402).
We also thank the assistants who helped
 with the observations for their hard work and kind cooperation.
\end{acknowledgements}

\end{document}